# Piston-assisted charge pumping

Davneet Kaur[1], Ilan Filonenko[2], and Lev Mourokh[1,3]

[1]*Department of Physics, Queens College of the City University of New York, Flushing, NY 11367, USA.*

[2]*Edgemont High School, Scarsdale, NY 10583, USA.*

[3]*The Graduate Center of CUNY, New York, NY 10016, USA*

We examine charge transport through a system of three sites connected in series in the situation when an oscillating charged piston modulates the energy of the middle site. We show that with an appropriate set of parameters, charge can be transferred against an applied voltage. In this scenario, when the oscillating piston shifts away from the middle site, the site's energy decreases and it is populated by a charge transferred from the lower energy site. On the other hand, when the piston returns to close proximity, the energy of the middle site increases and it is depopulated by the higher energy site. Thus through this process, the charge is pumped against the potential gradient. Our results can explain the process of proton pumping in one of the mitochondrial enzymes, Complex I. Moreover, this mechanism can be used for *electron* pumping in semiconductor nanostructures.

Proton-pumping complexes of mitochondrial membranes convert electron energy into the more stable form of a proton gradient across the membrane [1]. Though these complexes are well studied, the actual physical mechanisms of energy conversion have remained elusive in many cases. This is especially true for Complex I [2], where a distance of up to 15nm separates the electrons and protons. This enzyme consists of an L-shaped assembly of a hydrophobic arm embedded in the lipid membrane and a hydrophilic peripheral arm, which protrudes into the mitochondrial matrix. Electron transfers occur in the hydrophilic domain, while the proton pumps are located in the membrane domain.

The structure of the hydrophilic part was determined in Ref. [3], the membrane domain was resolved in Ref. [4], and the full structure was recently published in Ref. [5]. The membrane domain contains 4 proton pumps; each separated into two half-channels, see Fig.1. When there are no electrons in the hydrophilic domain (oxidized state), the residues at the end of the upper half-channels are assumed to be protonated by the negative side of the membrane, Fig 1(a). The appearance of an electron (reduction) leads to conformational changes, transferring the protons to the residues located at the beginnings of the lower half-channels. Subsequently, protons proceed to the positive side of the membrane, Fig 1(b).

The conformational changes that are induced by the electron transfer include the motion of a specific structure, helix HL. In the model proposed in Refs. [4,6], it was suggested that this motion might be responsible for proton pumping. In the oxidized state, this helix moves to the right to open the upper half-channels for protons (Fig. 1.) In the reduced state, it moves to the left to open the lower half-channels. This model does not, however, explain how the energy is transferred from the hydrophilic to the membrane domain in order to assist proton transfer against the population gradient. It is well established that in all electron-driven proton pumps, the electron-proton Coulomb interaction plays a crucial role in transferring energy from electron to proton degrees of freedom. While in Complex I, electrons and protons are well separated. Consequently, direct Coulomb coupling seems to be unlikely.

Here, we propose a model for the indirect electron-proton coupling, assisted by a piston (helix) with *non-uniform charge density*. This piston, with positive charges of magnitudes *e* accumulated at its edges, is located between the electron (to the right) and proton (to the left) sites. An electron populating the coupled site attracts such a piston, inducing a shift to the right. As a result, the piston moves away from the proton site, reducing its energy and facilitating its population. When the electron proceeds to the drain, the piston returns to its original position, increasing the energy of the *populated* proton site. Thus through this process, proton pumping occurs. It should be noted that the piston in our model could represent general conformation changes and not necessarily those of the helix HL as the crucial role of the helix in proton pumping was not confirmed in the mutation experiments [7].

In the present paper, we discuss a simplified model; see Fig. 2(a), in which the electron subsystem has been replaced by a periodic external force acting on the piston. This system consists of three proton sites (*A*, *B*, and *Q*) located between the source and drain, and the charged piston located near the middle site *Q*. The energies of the sites and the chemical potentials of the reservoirs are shown in Fig. 2(b). With the piston close to the site *Q*, its energy is greater than that of site *B*, while when the piston is moved away, it is lower than that of the site *A*. The interaction of the system with the environment (represented as a set of independent oscillators) leads (i) to its reorganization due to the charge transfers between the sites and (ii) to friction of the piston motion.

The Hamiltonian of this system is given by

$$H = E_A a_A^+ a_A + E_B a_B^+ a_B + E_Q a_Q^+ a_Q - \Delta_{AQ} a_Q^+ a_A - \Delta_{AQ}^* a_A^+ a_Q - \Delta_{QB} a_B^+ a_Q - \Delta_{QB}^* a_Q^+ a_B$$
$$+ \sum_k E_{Sk} s_k^+ s_k + \sum_k E_{Dk} d_k^+ d_k - \sum_k T_{Sk} a_A^+ s_k - \sum_k T_{Dk} a_B^+ d_k - \sum_k T_{Sk}^* s_k^+ a_A - \sum_k T_{Dk}^* d_k^+ a_B \quad , \quad (1)$$
$$+ \sum_j \frac{p_j^2}{2m_j} + \sum_j \frac{m_j \omega_j^2}{2} \left( x_j - C_{Aj} a_A^+ a_A - C_{Bj} a_B^+ a_B - C_{Qj} a_Q^+ a_Q \right)^2$$

where $a_\sigma^+ / a_\sigma$ are the proton creation/annihilation operators for the σ-site (σ = A,B,Q), $E_\sigma$ are the energies of these sites, $\Delta_{\sigma\sigma'}$ are the transfer amplitudes between the sites, $s_k^+ / s_k$ and $d_k^+ / d_k$ are the creation/annihilation operators for protons with wavevector $k$ for the source and drain, respectively, $T_{Sk}$ and $T_{Dk}$ are the transfer magnitudes between the sites and the reservoirs, $p_j$ and $x_j$ are the momentum and coordinate of the $j$-th harmonic oscillator with mass $m_j$ and frequency $\omega_j$, and $C_{\sigma j}$ are the coupling strengths of the proton-environment interaction.

The energy of the site $Q$ depends on the piston position $x$ as

$$E_Q = E_{Q0} + \frac{e^2}{4\pi\varepsilon\varepsilon_0} \frac{1}{\sqrt{(l_p + x)^2 + r_p^2}}, \quad (2)$$

where $l_p$ is the horizontal distance between the piston at equilibrium and the site $Q$ and $r_p$ is its vertical shift. We assume that the piston motion is in the overdamped regime. Thus, the piston's position obeys the phenomenological Langevin equation,

$$\zeta \frac{dx}{dt} = -kx + \frac{e^2}{4\pi\varepsilon\varepsilon_0} \frac{N_Q (l_p + x)}{\left( (l_p + x)^2 + r_p^2 \right)^{3/2}} + A(1 + \cos\Omega t) + \xi(t), \quad (3)$$

where $k$ is the elastic force costant, $N_Q$ is the population of the $Q$-site, $A$ and $\Omega$ are the amplitude and frequency, respectively, of the periodic force associated with the electron transfer in the hydrophilic domain, $\zeta$ is the drag coefficient, and $\xi$ is the fluctuation source (white noise) with zero mean value and the correlation function is given by

$$\langle \xi(t) \xi(t') \rangle = 2\zeta T \delta(t - t'). \quad (4)$$

Equations for the site populations can be derived using the equations of motion for the creation/annihilation operators of Eq. (1). It was shown previously [8-10] that in the high-temperature limit the resulting rate equations can be written as

$$\langle \dot{N}_A \rangle + \Gamma_S \langle N_A \rangle = \Gamma_S F_S(E_A) + \Phi_A$$
$$\langle \dot{N}_B \rangle + \Gamma_D \langle N_B \rangle = \Gamma_D F_D(E_B) + \Phi_B . \quad (5)$$
$$\langle \dot{N}_Q \rangle = -\Phi_A - \Phi_B$$

Here, angle brackets mean both quantum-mechanical and thermal averaging. $\Gamma_{S/D}$ are the coupling constants to the reservoirs given by

$$\Gamma_{S/D}(\omega) = \sum_k |T_{S/Dk}|^2 \delta(\omega - E_{S/Dk}) \quad (6)$$

and assumed to be frequency-independent. $F_{S/D}(E_{A/B})$ are the Fermi functions for the protons at the reservoirs,

$$F_S(E_A) = \frac{1}{\exp\{(E_A - \mu_S)/T\} + 1},$$

$$F_D(E_B) = \frac{1}{\exp\{(E_B - \mu_D)/T\} + 1},$$

(7)

where $\mu_{S/D}$ are the chemical potentials of the source/drain. Kinetic coefficients $\Phi_\alpha$ ($\alpha = A, B$) have the form,

$$\Phi_\alpha = \kappa_\alpha(E_\alpha - E_Q + \Lambda_\alpha)\langle N_Q \rangle\langle 1 - N_\alpha \rangle - \kappa_\alpha(E_Q - E_\alpha + \Lambda_\alpha)\langle N_\alpha \rangle\langle 1 - N_Q \rangle, \quad (8)$$

where

$$\kappa_\alpha(E) = |\Delta_{\alpha Q}|^2 \sqrt{\frac{\pi}{\Lambda_\alpha T}} \exp\left\{-\frac{E^2}{4\Lambda_\alpha T}\right\} \quad (9)$$

are the Marcus rates and

$$\Lambda_\alpha = \sum_j \frac{m_j \omega_j^2}{2}(C_{\alpha j} - C_{Qj})^2$$

(10)

are the reorganization energies of the environment due to the proton transfer.

The proton currents are given by

$$I_S = \frac{d}{dt}\sum s_k^+ s_k = \Gamma_S(N_A - F_S(E_A)),$$
$$I_D = \frac{d}{dt}\sum d_k^+ d_k = \Gamma_D(N_B - F_D(E_B)). \quad (11)$$

In the steady state regime, $I_S = - I_D$. The proton pumping occurs when the drain current is positive. So, protons are transferred from the reservoir with lower chemical potential to the reservoir with higher chemical potential.

Eqs. (4) and (5) are coupled, as the electrostatic force acting on the piston depends on the population of the middle site, while the energy of this site involved in Eqs. (8) and (9) depends on the piston position, as shown in Eq. (2). We solve these equations numerically, substituting the obtained values for the site populations in Eq. (11), and performing the time-averaging and averaging over possible realizations of the white noise $\xi$. The results are shown in Fig. 3 (a–c) for the following set of parameters: $\Omega = 10^9$ s$^{-1}$, $A = 41.4$ nN, $\Lambda_A = \Lambda_B = 50$ meV, $\Delta_{AQ} = \Delta_{BQ} = 25$ meV, $\Gamma_S = \Gamma_D = 10$ meV, $l_p = 0.5$ nm, $k = 8.9$ N m$^{-1}$, and $\zeta = 4.14$ nN m$^{-1}$ s (which corresponds to the diffusion coefficient $D = T/\zeta = 10^{-12}$ m$^2$ s$^{-1}$). The voltage applied across the membrane is 160 mV, so that the chemical potential of the source reservoir is assumed to be $\mu_s = -80$ meV and the chemical potential of the drain reservoir is $\mu_d = 80$ meV. The energy of the proton sites are $E_A = -150$ meV and $E_B = 250$ meV. These energies were chosen to prevent the back current at moderate temperatures, i.e., to ensure that the site $A$ is always populated from the source reservoir and depopulated by the site $Q$, and the proton is not transferred back to the reservoir. Correspondingly, the site $B$ is always depopulated by the drain reservoir, see the energy levels in Fig. 1(b).

The temperature dependence of the current is shown in Fig. 3(a) for a vertical separation of $r_p = 0.8$ nm between the site $Q$ and piston and the bare energy of the site $Q$ being $E_{Q0} = -200$ meV. It is evident from this figure that for the chosen set of parameters, the most effective operation of the proton pump occurs at physiological temperatures. At high temperatures the current becomes negative because the broadening of the Fermi functions of the reservoirs enables the back current which becomes dominant. The bare energy of $-200$ meV of the site $Q$ is optimal, as can be seen from Fig. 3(b). The dependence of the current on the vertical shift is presented in Fig. 3(c). From the figure, we can assess that if the piston and the site $Q$ are well separated, the electrostatic energy is not enough to raise the energy of site $Q$ above that of site $B$. Instead, the energies of the sites become close in value and this enhances the back current.

It is evident from our analysis that the proton pumping in mitochondria membranes can be achieved in the three-site system when the energy of the middle site is modulated by the moving piston. Similar system can be employed in semiconductor nanostructures to achieve *electron*

*pumping*. The energy scales in these systems are approximately 75 times smaller than that of the mitochondria, so the physiological-temperature phenomenology can be observed at 4 K. Electron pumping, the process when the net current is flowing between the unbiased electrodes (or even uphill), is usually achieved using surface acoustic waves (SAWs). SAWs are traditionally employed to investigate low-dimensional electron systems [11-13]. The electron pumping effect caused by SAWs was proposed theoretically in Ref. [14] and observed experimentally in Ref. [11]. We suggest that such pumping can occur without SAWs in the triple-quantum-dot system arranged in series between the leads, similar to the structure studied in Ref. [15]. The charged cantilever is placed in the vicinity of the middle quantum dot. Natural vibrations of the cantilever would change the separation between it and the quantum dot modulating the dot energy. With proper arrangement of the dot energy levels, the electron pumping can be achieved.

In summary, we examined possible piston-mediated mechanism of the proton pumping in Complex I of the mitochondria membranes. Three proton sites are placed between the source and drain reservoirs with chemical potential of the source (negative side of the membrane) being smaller than the chemical potential of the drain (positive side). The piston representing the electron-driven conformational changes of the actual complex has the positive charge near the edge which modulates the energy of the middle proton site. When the piston is far away, the energy of this site becomes smaller than the energy of the site near the source reservoir, so the middle site is populated. When the piston returns back, the energy of the middle site becomes larger than that of the site near the drain reservoir, and the proton is transferred there and eventually to the drain. Correspondingly, the proton pumping is achieved. We showed that for the set of parameters similar to the real system, the operation of our model is most effective at the physiological temperatures. We also suggested a possible implementation of this phenomenology in nanoelectronics. With the triple-quantum-dot system placed in series between the electrodes, the energy of the middle dot can be modulated using the charged cantilever to achieve uphill electron transport.

The work of D. K. and L. M. is partially supported by PSC-CUNY award 67080-00 45.

**Figures:**

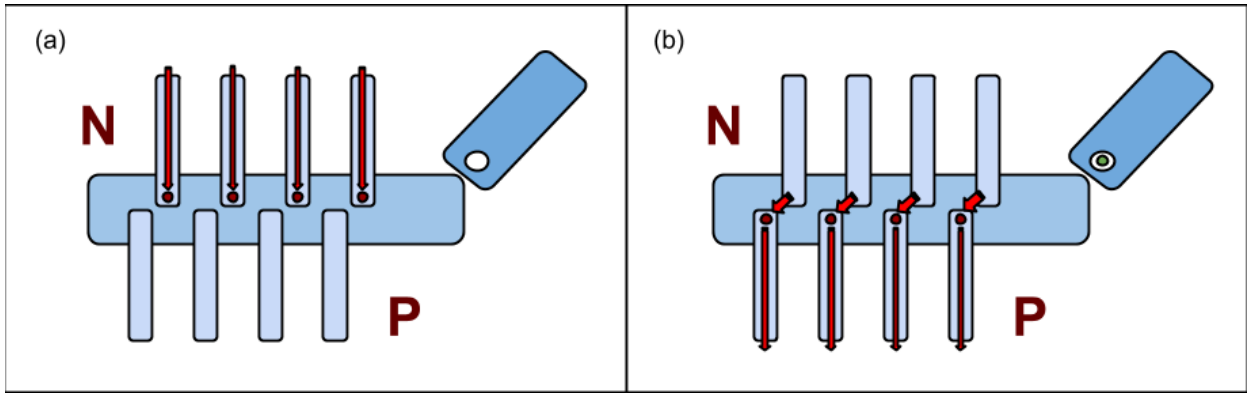

Fig. 1. Structure and operation of Complex I. (a) In the absence of an electron, the upper half-channels of the membrane domain are populated by protons from the negative side of the membrane. (b) Electron appearing in the hydrophilic domain leads to conformational changes moving protons to lower half-channels. Subsequently, protons are transferred to the positive side of the membrane.

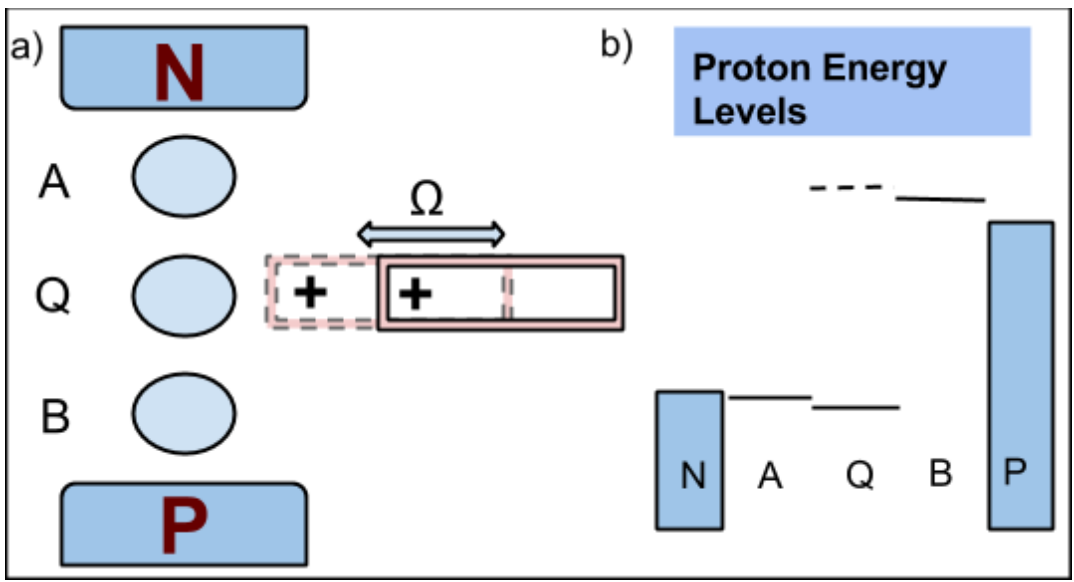

Fig.2. (a) Schematics of the model: Three proton sites are placed between the source and drain reservoirs. The piston having a charge near the edge oscillates in the vicinity of the middle site. (b) Energy diagram: The energy of the site $A$ is slightly below the chemical potential of the source, while the energy of the site $B$ is slightly above the chemical potential of the drain. The solid line represents the energy of the site $Q$ when the piston is moved far away from the site and the dashed line shows the energy of this site when the piston returns back.

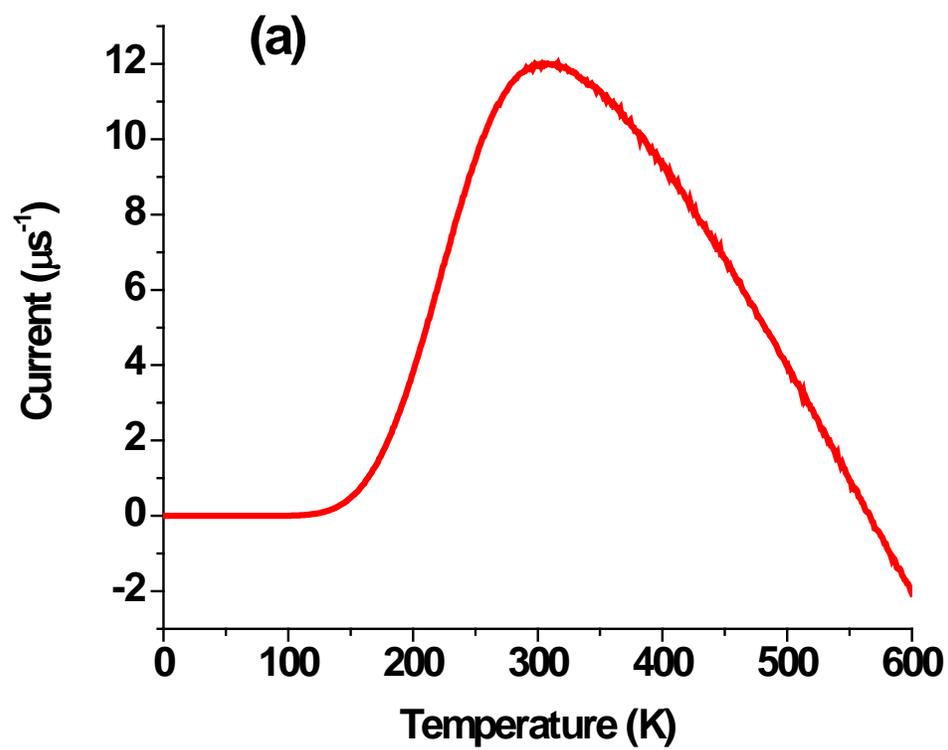

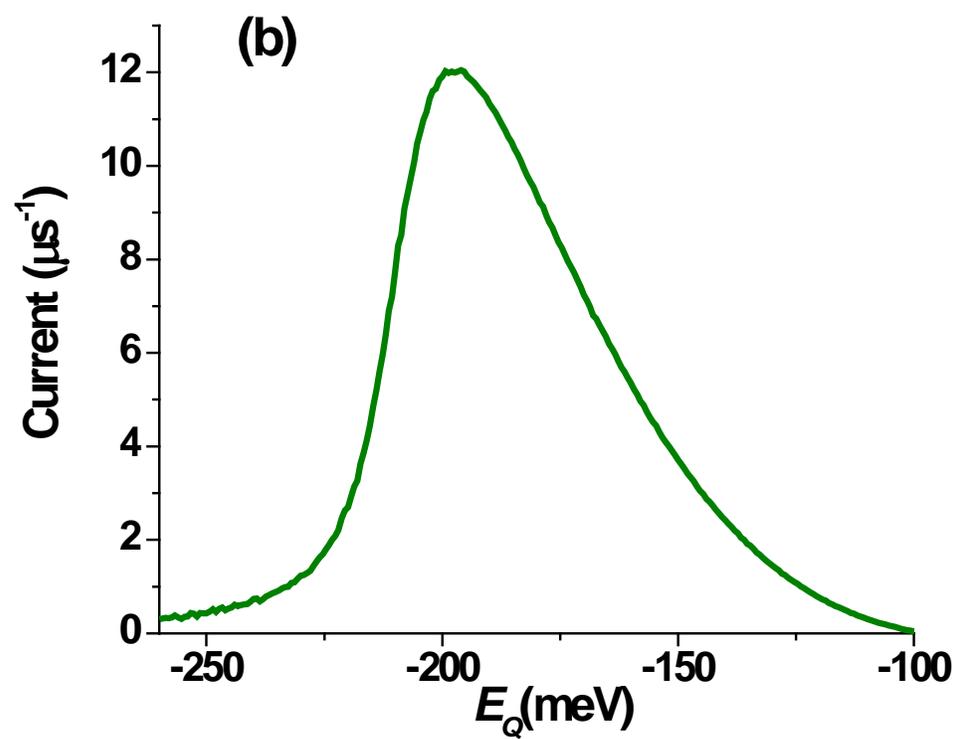

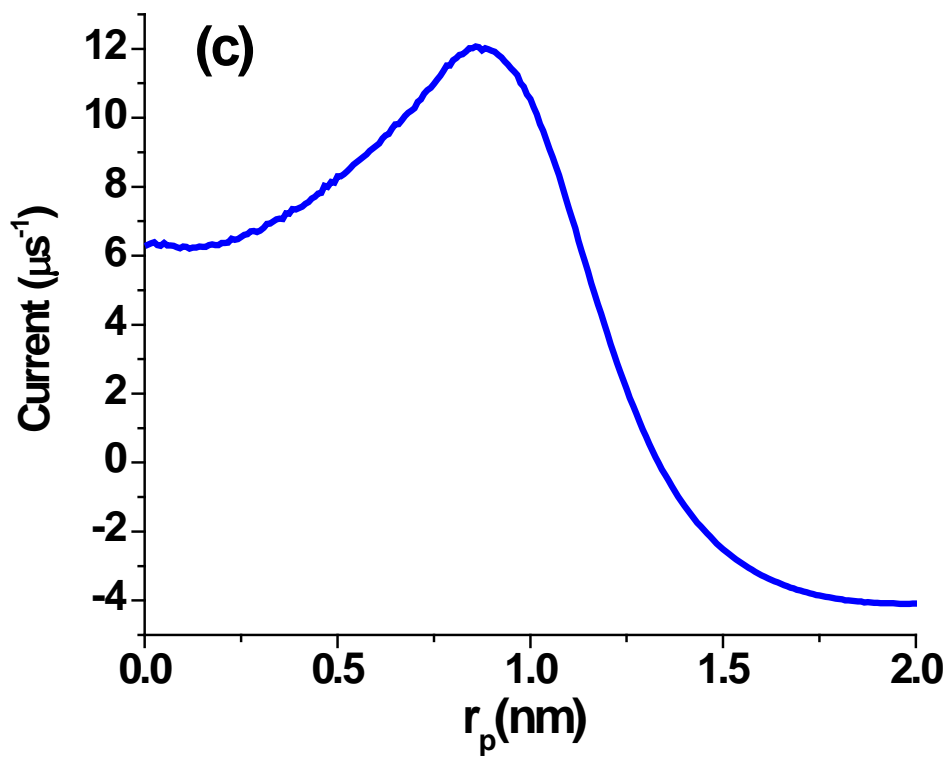

Fig. 3. (a) The temperature dependence of the proton current; (b) Dependence of the proton current on the unperturbed energy of the site $Q$; (c) Dependence of the proton current on the vertical shift of the piston with respect to the site $Q$.